\documentstyle[prl,aps]{revtex}
\begin{document}

\twocolumn[\hsize\textwidth\columnwidth\hsize
          \csname @twocolumnfalse\endcsname
\title{Unconventional phonon-mediated superconductivity in MgB$_{2}$}
\author{Guo-meng Zhao} 
\address{Department of Physics and Texas Center for Superconductivity, 
University of Houston, Houston, Texas 77204, USA\\}

\maketitle
\widetext

\begin{abstract}

We have evaluated the total carrier mass enhancement factor $f_{t}$ 
for MgB$_{2}$ from two independent experiments (specific heat and 
upper critical field). These experiments consistently show that $f_{t}$ = 
3.1$\pm$0.1. The unusually large $f_{t}$ is 
incompatible with the measured reduced gap (2$\Delta 
(0)/k_{B}T_{c}$ = 4.1) and the total 
isotope-effect exponent ($\alpha$ = 0.28$\pm$0.04) within the conventional 
phonon-mediated model. We propose an unconventional 
phonon-mediated mechanism, which is able to quantitatively explain the values of 
$T_{c}$, $f_{t}$, $\alpha$, and the reduced energy gap in a consistent 
way.

\end{abstract}
\vspace{1cm}
\narrowtext
]
The recent discovery of superconductivity near 40 K in MgB$_{2}$ 
\cite{Nagamatsu} has led to a remarkable excitement in
the solid-state physics community. Such a high-$T_{c}$ 
superconductivity in this simple intermetallic compound immediately 
raises a question of whether
mechanisms other than the conventional electron-phonon interaction are responsible for the superconductivity.  A significant 
boron isotope effect ($\alpha_{B} = -d\ln T_{c}/d\ln M_{B}$ = 
0.26$\pm$0.03, where $M_{B}$ is the mass of boron) 
\cite{Bud,Hinks,note} and 
nearly zero magnesium isotope effect ($\alpha_{Mg}$ = 0.02$\pm$0.01) \cite{Hinks} suggest that 
electron-phonon coupling plays an important role in the pairing 
mechanism. To explain the 40 K superconductivity, an electron-phonon coupling
constant of about 1 is needed.  First-principle calculations
give the coupling constant of 0.7-0.9 (Refs.~\cite{Kortus,Kong,Bohnen,Liu}), which appears to be sufficient to 
explain the observed high-$T_{c}$ superconductivity.  On the other 
hand, recent specific heat data 
\cite{Walti} indicate a very strong electron-phonon coupling with 
a total mass enhancement factor of about 3.2. This would suggest 
a coupling constant of 2.2 and 2$\Delta (0)/k_{B}T_{c} \approx$ 5 within the 
conventional phonon-mediated mechanism \cite{CarbotteRev}, in 
contradiction with the bulk-sensitive Raman scattering experiments 
which show 2$\Delta (0)/k_{B}T_{c}$ = 4.1 (Ref.\cite{Chen}).  

Here we evaluate the carrier mass enhancement factor $f_{t}$ from the measured 
upper critical field. The deduced 
$f_{t}$ is 3.1$\pm$0.1, in remarkably good agreement with that deduced from 
the independent specific heat data \cite{Walti}. The large mass enhancement factor is 
not compatible with the measured gap amplitude and the reduced total 
isotope-effect exponent within the conventional 
phonon-mediated model. We thus propose an unconventional 
phonon-mediated mechanism where both the long-range Fr\"ohlich-type 
and short-range electron-lattice interactions are considered and 
treated distinctively.  Within this 
scenario, we are able to quantitatively explain the $T_{c}$ value, the 
mass enhancement factor, the 
total isotope-effect exponent, and the reduced energy gap.

For a clean superconductor, as it is the case for MgB$_{2}$ 
\cite{Canfield,Bud1}, the 
zero-temperature coherence length $\xi (0)$ is equal to the BCS coherence 
length \cite{CarbotteRev}, that is,
\begin{equation}
\xi (0) = \frac{\hbar v_{F}}{\pi\Delta (0)}.
\end{equation}
Here $v_{F}$ is the renormalized Fermi velocity, i.e., $v_{F}$ = 
$v^{b}_{F}/f_{t}$ ($v^{b}_{F}$ is the bare Fermi velocity). Then
\begin{equation}\label{eq2}
\xi (0) = \frac{\hbar v^{b}_{F}}{f_{t}\pi\Delta (0)}.
\end{equation} 

Using  $H_{c2}(0) = \Phi_{\circ}/2\pi[\xi(0)]^{2}$ and $H_{c2}(0)$ = 
15$\pm$2 T (Ref. \cite{Bud1}), we obtain $\xi (0)$ = 47$\pm$2 \AA. With $v^{b}_{F}$ = 
4.85$\times$10$^{5}$ m/s (Ref.~\cite{Kortus,note2}), $\Delta (0)$ = 6.9 meV 
\cite{Chen}, and $\xi (0)$ = 47$\pm$2 \AA, 
we find from Eq.~\ref{eq2} that $f_{t}$ = 3.1$\pm$0.1.  Remarkably, the 
deduced mass enhancement factor $f_{t}$ here is in excellent agreement 
with that found from specific heat data \cite{Walti}.

It is clear that the large mass enhancement factor is not compatible 
with the measured energy gap (2$\Delta (0)/k_{B}T_{c}$ = 4.1) 
\cite{Chen} and the reduced isotope 
exponent ($\alpha$ = 0.28$\pm$0.04) \cite{note} within the conventional phonon-mediated model.  In order to resolve 
the above controversy, one needs to consider a long-range 
Fr\"ohlich-type electron-phonon interaction that results from the 
Coulomb interactions of electronic charge carriers with the ions of 
ionic (polar) materials \cite{Emin}.  This interaction is distinct from the 
short-range electron-lattice interaction (e.g., deformation-potential interaction, 
Holstein interaction) that has a carrier's energy only depending on 
the positions of the atoms with which it overlaps. The Fr\"ohlich 
electron-phonon interaction is particularly strong in an ionic solid 
or a perovskite oxide where its static dielectric constant 
$\epsilon_{\circ}$ is always much 
larger than its high-frequency dielectric constant $\epsilon_{\infty}$, whereas the 
short-range electron-lattice interaction (including the 
electron-lattice interaction with acoustic phonons) is present in all 
the materials \cite{Emin}.   Recent Quantum Monto Carlo simulation 
\cite{Alex} has 
shown that the Fr\"ohlich-type electron-phonon interaction is 
nonretarded and can always lead to a mass 
enhancement factor of $f_{p} = \exp (g^{2})$ no matter how weak this 
interaction is. Here $g^{2} = A/\omega_{\circ}$, $A$ 
is a constant, and $\omega_{\circ}$ is the characteristic frequency of 
optical phonons. It is apparent that this mass enhancement factor 
$f_{p}$ will 
strongly depend on the 
isotope mass if $g^{2}$ is substantial, in contrast to the 
conventional retarded electron-phonon coupling model where the mass 
enhancement factor is essentially isotope-mass independent 
\cite{CarbotteRev}. On the other hand, the short-range component of the 
electron-lattice interaction is retarded and can be 
modeled within the conventional Eliashberg equations when the coupling constant is 
less than 1 and the phonon energies are much lower than the bare 
hopping integral \cite{Alex,ale}.  When only a short-range 
electron-lattice interaction is present and the coupling constant is 
far larger than 1, the interaction becomes nonretarded and small 
polarons can be formed \cite{Emin,ale}. In this case, the Migdal adiabatic 
approximation 
breaks down \cite{ale}  and one cannot use the Eliashberg equations to 
describe superconductivity \cite{Emin,ale}.

Since the Fr\"ohlich-type electron-phonon 
interaction is nonretarded, we can assume that the role of this 
interaction is simply to enhance the density of states \cite{ale} and reduce the 
direct Coulomb interaction 
between two carriers \cite{Emin,ale}.  In other words, we can model the 
retarded short-range electron-phonon 
coupling within  
the conventional Eliashberg equations, but the effective density of states 
and the coupling constant $\lambda$ are enhanced by a factor of 
$f_{p}$ due to the presence of the Fr\"ohlich-type electron-phonon 
interaction. The effective Coulomb 
pseudopotential $\mu^{*}$ in the Eliashberg equations will not change significantly 
when the Fr\"ohlich electron-phonon 
interaction sets in. This is because this interaction enhances the density of 
states, which tends to increase $\mu^{*}$, while it produces an 
attractive potential that effectively reduces the direct Coulombic 
repulsion by a factor of about $\epsilon_{\circ}/\epsilon_{\infty}$ 
\cite{Emin} and thus tends to decrease $\mu^{*}$ (see Eq.~4 below).  Therefore, the Fr\"ohlich-type electron-phonon interaction can 
enhance superconductivity mainly through increasing the effective coupling 
constant for the retarded electron-phonon interaction. For ionic materials (e.g., 
MgB$_{2}$),  this type of electron-phonon coupling should be rather strong, 
so that the superconductivity can be enhanced substantially.

Within this simplified approach, the effective coupling constant for the 
retarded short-range electron-phonon interaction is $\lambda = \lambda_{b}f_{p}$, 
where $\lambda_{b}$ is the bare short-range retarded electron-phonon coupling 
constant in 
the absence of the Fr\"ohlich-type electron-phonon interaction. The 
value of $\lambda_{b}$ for MgB$_{2}$ has been calculated to be 
0.7-0.9 (Ref.\cite{Kortus,Kong,Bohnen,Liu}). The 
total carrier mass enhancement factor is then given by
\begin{equation}\label{ft}
f_{t} = f_{p}(1 + \lambda_{b}f_{p}).
\end{equation}
The factor $1 + \lambda_{b}f_{p}$ is the mass enhancement factor due 
to the retarded electron-phonon interaction with the enhanced 
coupling constant $\lambda = \lambda_{b}f_{p}$. The Coulomb pseudopotential $\mu^{*}$ is
\begin{equation}
\mu^{*} = \frac{\mu}{1 +\mu\ln (E_{F}/\hbar\omega_{ln})},
\end{equation}
Here $\mu \propto Uf_{p}$ ($U$ is the effective Coulomb interaction 
between two carriers, which is renormalized by the Fr\"ohlich 
interaction ); $E_{F} = E^{b}_{F}/f_{p}$ ($E^{b}_{F}$ is 
the bare Fermi energy); $\omega_{ln}$ is the logarithmically averaged frequency, which 
is normally lower than the Debye frequency $\omega_{D}$ by a factor of 
about 1.2. Indeed, the calculated $\hbar\omega_{ln}$ is 53.8 meV 
(Ref.~\cite{Osborn}), which 
is a factor of 1.2 smaller than the measured $\hbar\omega_{D}$ (64.3 
meV) \cite{Bud,Walti}.

Since  $f_{p} = \exp (g^{2})$, which depends on the isotope mass 
\cite{Alex}, the 
quantities $\lambda$, $\mu^{*}$, and $f_{t}$ are all isotope-mass dependent. 
One can easily deduce that the total exponents 
of the isotope 
effects on $\lambda$ ($\alpha_{\lambda} = -\sum d\ln \lambda /d\ln 
M_{j}$, where $M_{j}$ is the mass of the $j$th atom in the unit 
cell) and on $\mu^{*}$ ($\alpha_{\lambda} = -\sum d\ln \mu^{*}/d\ln 
M_{j}$) are given by
\begin{equation}
\alpha_{\lambda} = - \frac{1}{2}\ln f_{p},
\end{equation}
and 
\begin{equation}
\alpha_{\mu^{*}} = \alpha_{\lambda}[1 - \mu^{*}\ln 
(E_{F}/\hbar\omega_{ln})] + \mu^{*}(\alpha_{\lambda} + \frac{1}{2}).
\end{equation}
Furthermore, from the McMillian formula \cite{CarbotteRev},
\begin{equation}
k_{B}T_{c} = \frac{\hbar\omega_{ln}}{1.2}\exp [-\frac{1.04(1 + 
\lambda)}{\lambda - \mu^{*}(1 + 0.62\lambda)}],
\end{equation}
we can also determine the total exponent of the isotope effect on 
$T_{c}$ ($\alpha = -\sum d\ln T_{c}/d\ln 
M_{j}$),
\begin{eqnarray}
\alpha =  \frac{1}{2} + \frac{1.04(1 + 0.38\mu^{*})\lambda}
{[\lambda - \mu^{*}(1 + 0.62\lambda)]^{2}}\alpha_{\lambda} \nonumber 
\\
~-\frac{1.04(1 + \lambda)(1 + 0.62\lambda)\mu^{*}}{[\lambda - \mu^{*}(1 + 
0.62\lambda)]^{2}}\alpha_{\mu^{*}}.
\end{eqnarray}

From the above equations, we can calculate $f_{p}$, $\lambda$, 
$\hbar\omega_{ln}$, $\alpha_{\lambda}$, $\alpha_{\mu^{*}}$, and $\alpha$ as a function of 
$\lambda_{b}$ using fixed parameters $f_{t}$ = 3.2, $T_{c}$ = 40 K, 
$E_{F}$ = 0.57 eV (Ref.~\cite{Lorenz}), and $\mu^{*}$ = 0.1. In Fig.~1, 
we plot the calculated $\alpha$, 
$\hbar\omega_{ln}$, and $f_{p}$ as a function of $\lambda_{b}$. One can see that the 
calculated $\alpha$ is 0.3 for $\lambda_{b}$ = 0.8, a value 
lying within the first-principle calculations ($\lambda_{b}$ = 
0.7-0.9) \cite{Kortus,Kong,Bohnen,Liu}. 
The calculated isotope exponent $\alpha$ is in quantitative agreement 
with the measured one (0.28$\pm$0.04) \cite{note}. The reduction in 
the isotope 
exponent is due to the fact that the coupling constant $\lambda$ has a 
negative isotope effect which partially cancels out the positive isotope effect on 
the prefactor of Eq.~7. As $\lambda_{b}$ increases, $f_{p}$ must 
decrease to keep $f_{t}$ a constant (see Eq.~3 and Fig.~1c). The decrease of $f_{p}$ reduces the magnitude of the negative isotope 
effect on the coupling constant, and thus increase the isotope effect 
on $T_{c}$, as seen clearly in Fig.~1a.

Meanwhile, the  calculated $\hbar\omega_{ln}$ is about 40 meV for $\lambda_{b}$ = 0.8 (see 
Fig.~1b), which 
appears to be lower than the one (53.8 meV) determined from the phonon 
density of states \cite{Osborn}. However, inelastic neutron 
scattering experiments  \cite{Sato} show that a low-energy phonon mode at about 17 meV 
is strongly coupled to conduction electrons; the intensity of the 17 
meV peak increases with decreasing temperature
for $T > T_{c}$ whereas below $T_{c}$ it starts decreasing. The 
strong coupling to the low 
energy mode is not expected from the theoretical calculations 
\cite{Kortus,Kong,Bohnen,Liu}, and may be related to some kind of structural 
instability \cite{Sato}.  If this is true,  the
calculated  $\hbar\omega_{ln}$ in Ref.~\cite{Osborn} should be substantially 
overestimated. Indeed, from the measured 2$\Delta (0)/k_{B}T_{c}$ = 4.1, one can 
determine the magnitude of $\hbar\omega_{ln}$ using a formula 
\cite{CarbotteRev}
\begin{equation}\label{gap}
\frac{2\Delta (0)}{k_{B}T_{c}} = 3.53[1 + 
12.5(\frac{k_{B}T_{c}}{\hbar\omega_{ln}})^{2}\ln(\frac{\hbar\omega_{ln}}{2k_{B}T_{c}})].
\end{equation}
Substituting  $T_{c}$ = 40 K and 2$\Delta (0)/k_{B}T_{c}$ = 4.1 into 
Eq.~\ref{gap}, we get $\hbar\omega_{ln}$ = 40 meV, in quantitative 
agreement with the above independent calculation.

From Fig.~1c, one can see that $f_{p}$ is about 1.5 at $\lambda_{b}$ = 
0.8, that is, the Fr\"ohlich-type electron-phonon 
interaction enhances the density of states and the coupling constant by a factor of 1.5. 
Without this interaction, the material would have a transition 
temperature of about 22 K. Therefore, the Fr\"ohlich-type electron-phonon 
interaction in ionic materials can indeed enhance superconductivity.

Although the conventional phonon-mediated model could account for the 
observed $T_{c}$ value \cite{Kortus,Kong,Bohnen,Liu}, it cannot 
consistently explain the $T_{c}$ value, the reduced isotope exponent ($\approx$0.3), the 
large carrier mass enhancement factor (3.1), and the reduced energy 
gap (2$\Delta (0)/k_{B}T_{c}$ = 4.1). The accurate determination of 2$\Delta 
(0)/k_{B}T_{c}$ = 4.1 places a strong constraint on $\hbar\omega_{ln}$, that 
is, $\hbar\omega_{ln}$ $\simeq$ 40 meV. With $\hbar\omega_{ln}$ = 40 meV, and $T_{c}$ = 
40 K, one cannot find any parameters that could lead to $\alpha$ 
$\simeq$ 0.28.  Further, 
one would never get a carrier mass enhancement factor 
larger than 3 within the conventional model. On the other hand, the 
proposed unconventional phonon-mediated mechanism 
naturally resolves these controversies and is able to quantitatively explain 
these experiments in a consistent way. In addition, an important 
prediction of this scenario is that both electronic specific heat 
$\gamma$ and 
London penetration depth $\lambda_{L}(0)$ depend on the isotope mass of boron. One can 
easily show that, upon replacing $^{10}$B with $^{11}$B, both $\gamma$ 
and $\lambda^{2}_{L}(0)$
will decrease by about 4$\%$. The isotope effect on $\gamma$ should be 
observable if one could measure the specific heat of the isotope samples 
down to a low temperature ($<$ 1 K) under a magnetic field higher 
than $H_{c2}(0)$. On the other hand, a special cation must be taken in 
determination of the 
intrinsic London penetration depth since the extrinsic contribution to the 
measured penetration depth due to defects \cite{Zhaosymmetry} would mimic the 
isotope effect if two isotope samples have different densities of defects.

In summary, the carrier mass enhancement factor $f_{t}$ has been  
determined for MgB$_{2}$ from 
the measured upper critical field and the calculated bare Fermi velocity. It is found 
that $f_{t}$ is 3.1$\pm$0.1, in remarkably good agreement with that deduced from 
the independent specific heat data. The unusually large $f_{t}$ is 
inconsistent with the measured reduced gap and the total 
isotope-effect exponent ($\alpha \approx$ 0.3) within the conventional 
phonon-mediated model. We thus propose an unconventional 
phonon-mediated mechanism where long-range Fr\"ohlich 
electron-phonon interaction and short-range retarded 
electron-phonon interaction are modeled separately. Within this 
scenario, we are able to quantitatively explain the values of 
$T_{c}$, $f_{t}$, $\alpha$ and 2$\Delta 
(0)/k_{B}T_{c}$. 
~\\
~\\
{\bf Acknowledgment:} The work at Houston was supported in 
part by the US National Science Foundation, and work at Berkeley by the 
US Department of Energy.

\begin{figure}[htb]
\input{epsf}
\epsfxsize 7cm
\centerline{\epsfbox{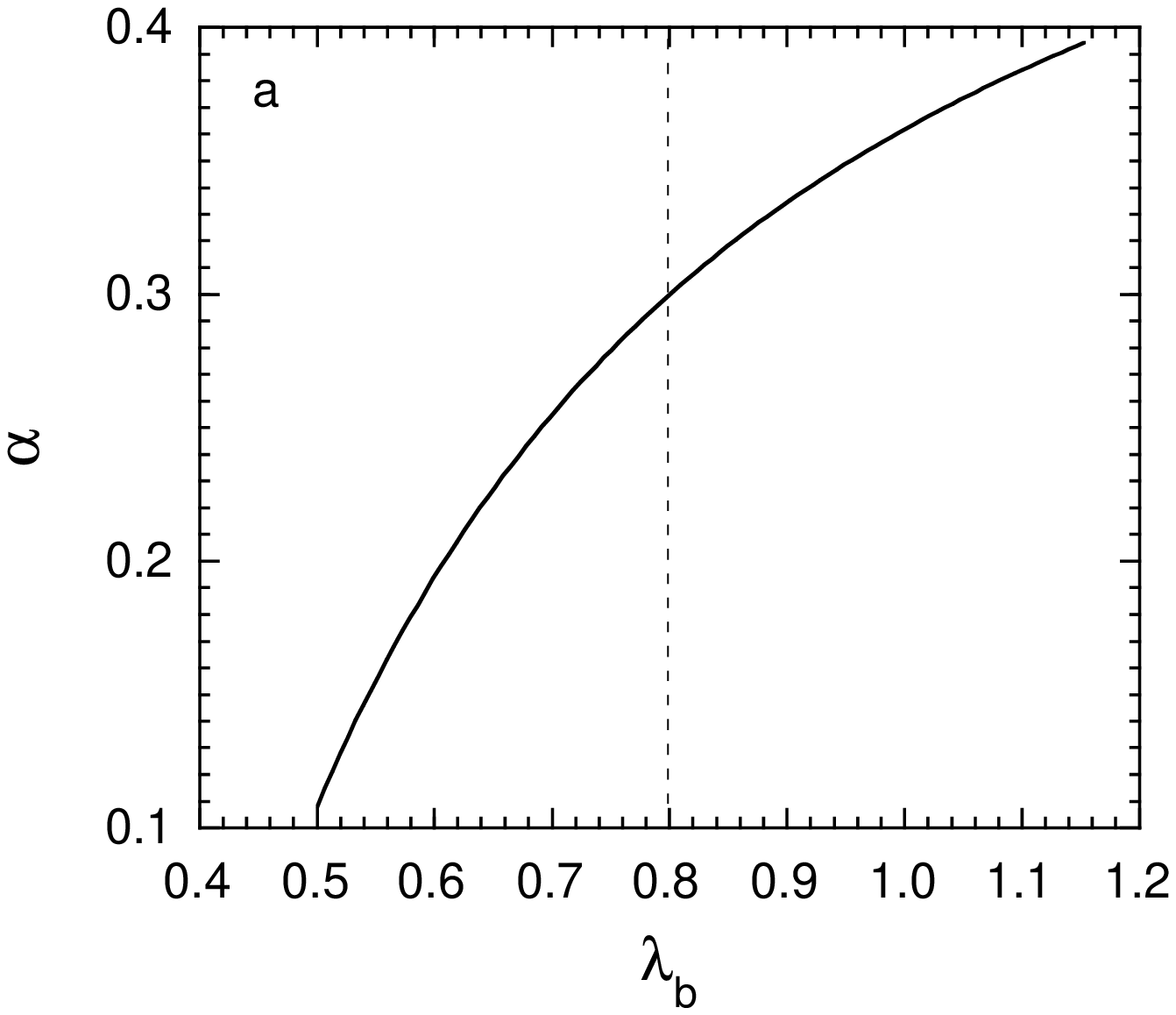}}
\vspace{0.4cm}
\input{epsf}
\epsfxsize 7cm
\centerline{\epsfbox{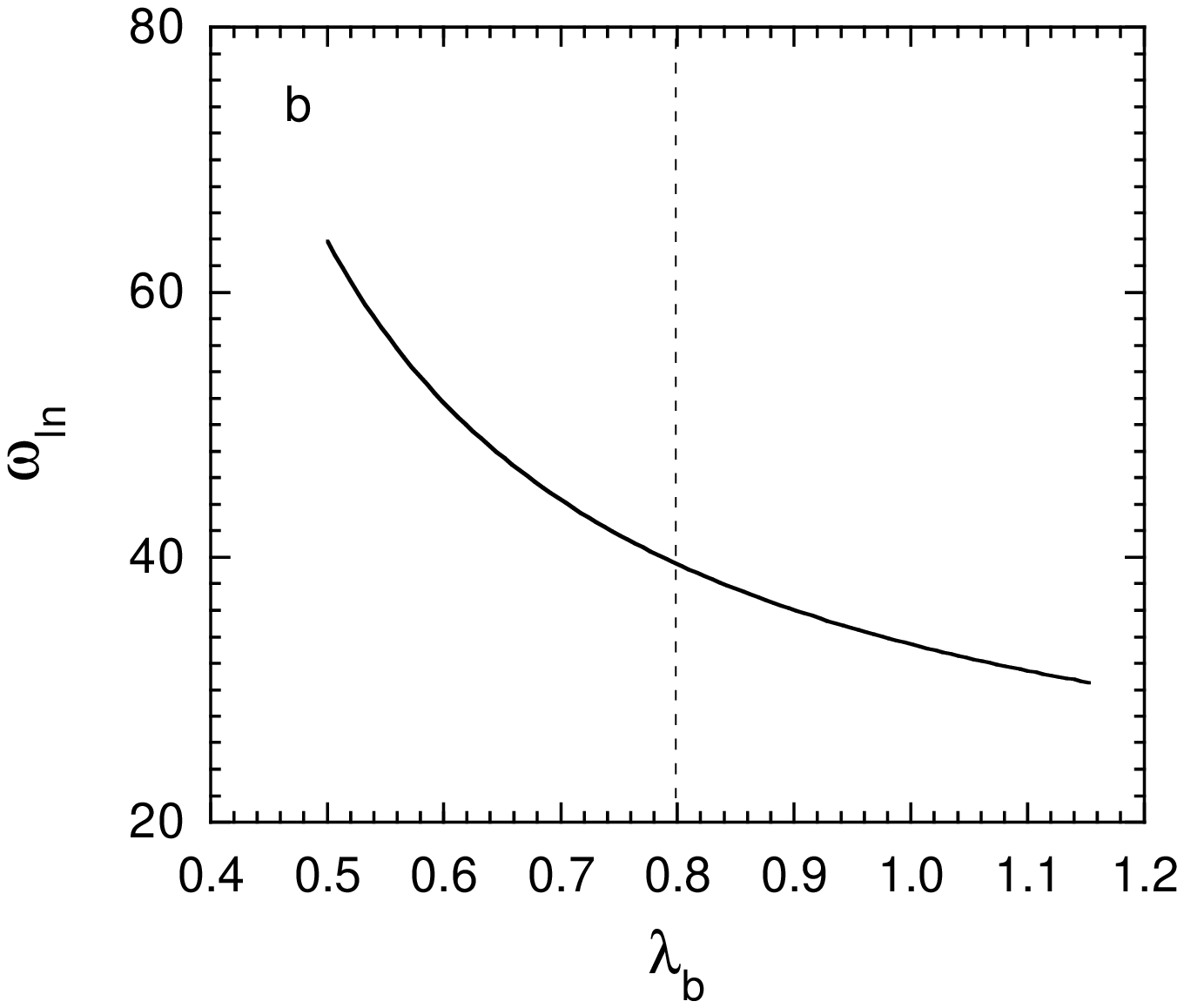}}
\vspace{0.4cm}
\input{epsf}
\epsfxsize 7cm
\centerline{\epsfbox{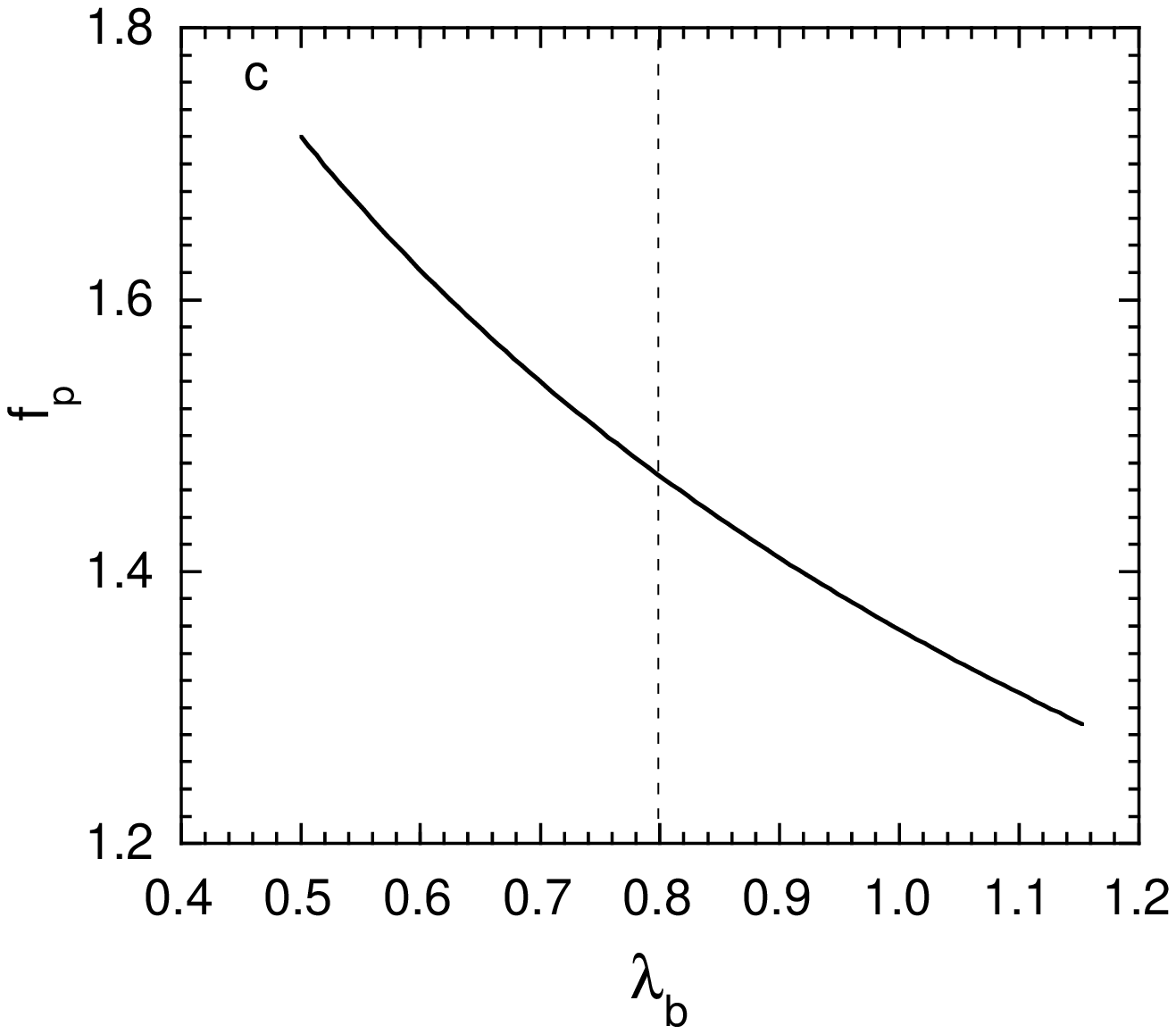}}
    \vspace{0.3cm}
	\caption[~]{The total isotope exponent $\alpha$, 
the logarithmically averaged frequency $\hbar\omega_{ln}$, and 
the mass enhancement factor $f_{p}$ (due to the Fr\"ohlich 
electron-phonon interaction)  as a function of the bare 
retarded electron-phonon coupling constant $\lambda_{b}$.}
\end{figure}


\begin{thebibliography}{99}
\bibitem{note}The measured boron isotope shift is 1.0$\pm$0.1 K 
\cite{Bud} and 
1.12$\pm$0.05 K \cite{Hinks}. Using our definition $\alpha_{B} = -d\ln 
T_{c}/d\ln M_{B}$, we evaluate $\alpha_{B}$ = 0.26$\pm$0.03. With 
$\alpha_{Mg}$ = 0.02$\pm$0.01 (Ref.~\cite{Hinks}), we obtain the total 
isotope exponent $\alpha$ = 0.28$\pm$0.04.
\bibitem{note2}The bare Fermi velocity $v_{F}^{b}$ is a simple 
geometric average of the calculated Fermi velocities along three axes 
\cite{Kortus}.
\end{thebibliography}
\end{document}